\documentclass{pnastwo}

\usepackage[dvips]{graphicx}
\usepackage{pnastwof}
\usepackage{amssymb,amsfonts,amsmath}

\def\be{\begin{equation}}
\def\ee{\end{equation}}
\newcommand{\ba}{\begin{eqnarray}}
\newcommand{\ea}{\end{eqnarray}}
\newcommand{\rd}{\mathrm{d}}            % Roman d for differential
\newcommand{\re}{\mathrm{e}}            % Roman e for exponential
\newcommand{\ri}{\mathrm{i}}
\newcommand{\rIm}{\mathrm{Im}}
\newcommand{\rRe}{\mathrm{Re}}

\newcommand{\te}[1]{\mbox{\boldmath $ #1 $}}

\newcommand{\bn}{\te{n}}
\newcommand{\bs}{\te{s}}
\newcommand{\bbe}{\te{e}}
\newcommand{\bu}{\te{u}}
\newcommand{\bv}{\te{v}}

\newcommand{\bF}{\te{F}}
\newcommand{\bU}{\te{U}}
\newcommand{\bE}{\te{E}}

\begin{document}

%\conflictofinterest{Conflict of interest footnote placeholder}

%\track{Insert 'This paper was submitted directly to the PNAS office.' when applicable.}

%\footcomment{Abbreviations: SAM, self-assembled monolayer; OTS, octadecyltrichlorosilane}

\title{A frictionless microswimmer}

\author{Alexander M. Leshansky\affil{1}{Department of Chemical Engineering},
Oded Kenneth\affil{2}{Department of Physics, Technion-IIT, Haifa 32000, Israel},
Omri Gat\affil{3}{Racah Institute of Physics, Hebrew University, Jerusalem 91904, Israel},
 \and
Joseph E. Avron\affil{2}{}\thanks{To whom correspondence should be addressed. E-mail: \texttt{avron@tx.technion.ac.il}}}

\contributor{Submitted to Proceedings of the National Academy of Sciences of the United States of America}

\maketitle

\begin{article}

\begin{abstract} We investigate the self-locomotion of an elongated microswimmer by virtue of the unidirectional tangential surface treadmilling. We show that the propulsion could be almost frictionless, as the microswimmer is propelled forward with the speed of the backward surface motion, \textit{i.e.} it moves throughout an almost quiescent fluid. We investigate this swimming technique using the special spheroidal coordinates and also find an explicit closed-form optimal solution for a two-dimensional treadmiler via complex-variable techniques.

\end{abstract}


%%%%%%%%%%%%%%%%%%%%%%%%%%%5

\dropcap{T}iny swimmers, be they micro-organisms or microbots, live in a world dominated by friction \cite{purcell}. In this world, technically, the world of low Reynolds numbers, motion is associated with energy dissipation. In the absence of external energy supply objects rapidly come to rest \cite{berg}. It is both conceptually interesting, and technologically important, to try and understand what classes of strategies lead to effective swimming in a setting dominated by dissipation. A particularly promising class of strategies is where the motion is, in a sense, only apparent; where a shape moves with little or no motion of material particles.

The wheel is the mechanical application of this strategy and it is instructive to examine it from this point of view. The (unholonomic) constraint of rolling without slipping comes about because of the large friction between the wheel and the surface supporting it. Nevertheless, a wheel can roll with little or no dissipation of energy. One way to view this is to note that the motion of the point of contact is only apparent. The point of contact moves, even though the part of the wheel in contact with the surface,  is momentarily at rest.

An example closer to the world of low Reynolds numbers, is the actin-based propulsion of the leading edge of motile cells \cite{PCC01}, intracellular bacterial pathogens \cite{TP89} and biomimetic cargoes \cite{CFOT99,Anne02,UCASO03}. The actin filaments assemble themselves from the ambient solution at the front end and disassemble themselves at the rear end. Here again, it is only shape that is moving and in principle at least the energy invested at the front end can be recovered at the rear end. (There are thermodynamics and entropic issues that we shall not consider here.) Apparent motions seem to be good at fighting dissipation \cite{Leshansky06}.

Here we shall focus on a closely related mode of locomotion: surface treadmilling. In surface treadmilling the swimmer moves without a change of shape, by a tangential surface motion. Surface is generated at the front end and is consumed at the rear end\footnote{Alternatively, we may think of a a slender microbot that is topologically equivalent to a toroidal swimmer proposed by Purcell \cite{purcell}, \textit{i.e.} the surface is not created or destroyed, but rather undergoes a continuous tank-treading motion}. In contrast to actin and microtubules, the surface treadmilling does not rely on the exchange of material with the ambient fluid. (The swimmer needs, of course,  an inner mechanism to transfer material from its rear to its front). It is intuitively clear that a needle shaped swimmer undergoing treadmilling can move with very little dissipation because the ambient fluid is almost quiescent and there is almost no relative motion between the surface of the swimmer and the fluid. One can not make treadmilling completely non-dissipative because there is always some remanent dissipation associated with the motion of the front and rear ends. The main question that we shall address here is how can one quantitatively estimate this remanent dissipation.

Let us first consider simple qualitative estimates of the power dissipated by viscosity in treadmilling. Consider, a rod-like slender swimmer of length $\ell$, thickness $d$  with rounded caps undergoing treadmilling at velocity $U$. It is reasonable to assume (and the analysis in the following section can be used to justify) that all the dissipation is associated with the rounded ends. Hence, by dimensional analysis, the power dissipated in treadmilling is of the order of $\mu d U^2$ where $\mu$ is the viscosity coefficient. Let us compare  this with the power needed to drag the ``frozen" treadmiler. By Cox slender body theory \cite{KK91} the power needed to drag the tube with velocity $U$ is $\sim \mu \ell U^2/\log{(2 \ell/d)}$. Hence the ratio of power invested in dragging and swimming scales like $(\varepsilon \log{1/\varepsilon})^{-1}$ and can be made arbitrarily large. Here $\varepsilon=d/\ell$ is the aspect ratio of the swimmer.

One can now ask if there are slender treadmilers that are arbitrarily better than the slender rod-like treadmiler above? Consider now an elongated ellipsoidal microswimmer whose surface is given by $z^2/b^2+r^2/a^2=1$ where $r^2\equiv x^2+y^2$. Let us assume again, that the viscous dissipation is a result of a tip propulsion, and
estimate the position of the tip from the condition $|\rd r/\rd z|=1$.
%the position of the tip can be determined approximately from the condition $|\rd r/\rd z|\leq1$.
%(where the angle between the normal to the surface and the propulsion direction is 45$^\circ$).
It can be readily demonstrated that in the case of $\varepsilon=a/b\ll 1$ the tip is located at a distance of $b(1-{1\over2}\varepsilon^2)$
from the center and its typical width scales as $ a\varepsilon$. Therefore, applying the same arguments as before, the dissipation rate should then scale as $\mathcal{P}\sim \mu a \varepsilon U^2$ and the ratio of power expanded in dragging and swimming becomes $\frac{b \mu U^2/\log{(2 b/a)}}{\mu a\varepsilon U^2}\sim\frac{1}{\varepsilon^2 \log{1/\varepsilon}}$.

We shall see, by a more accurate analysis, that for prolate spheroid the ratio of power in treadmilling to dragging is of the order of $(\varepsilon\log{\varepsilon})^{-2}$. In the following sections we shall analyze two models of treadmillers in 3 and 2 dimensions, respectively.

\subsection{The theoretical framework} We model the micro-swimmer as a prolate spheroid swimming in an unbounded fluid by continuous tangential surface motion.  The  Cartesian-coordinate system $(x_1,x_2,x_3)$ is fixed with the center ${\it O}$ of the spheroid. A modified orthogonal prolate spheroidal coordinate system $(\tau,\zeta,\varphi)$ with unit vectors $(\bbe_\tau, \bbe_\zeta,\bbe_\varphi)$ is defined via the relations $x_1=c\{\tau^2-1\}^{1/2}\{1-\zeta^2\}^{1/2}\cos\varphi$, $x_2=c\{\tau^2-1\}^{1/2}\{1-\zeta^2\}^{1/2}\sin\varphi$ and $x_3=c\tau\zeta$,
where $-1\le\zeta\le1\le\tau<\infty$, $0\le\varphi\le 2\pi$ and $c$ is the semi-focal distance \cite{DHP94}. The coordinate surfaces $\tau=\tau_a=const$ are a family of confocal spheroids, $x_3^2/b^2+(x_1^2+x_2^2)/a^2=1$, centered at the origin with major and minor semi-axis given by $b=c\tau_a$ and $a=c\{\tau_a^2-1\}^{1/2}$, respectively.
%The Lam\'e metric coefficients are
%\[
%\widehat{H}_{\tau}=c\frac{\{\tau^2-\zeta^2\}^\frac{1}{2}}{\{\tau^2-1\}^\frac{1}{2}}\:, %\widehat{H}_{\zeta}=c\frac{\{\tau^2-\zeta^2\}^\frac{1}{2}}{\{1-\zeta^2\}^\frac{1}{2}}\:, %\widehat{H}_{\varphi}=c\{\tau^2-1\}^\frac{1}{2}\{1-\zeta^2\}^\frac{1}{2}.
%\]
We assume that a steady axisymmetric flow has been established around the micro-swimmer as a result of the tangential surface treadmilling with a uniform far-field velocity $U$ (equal to the laboratory frame propulsion speed) in the negative $x_3$-direction. The low-Re incompressible flow is governed by the Stokes and continuity equations,
\be
\Delta \bv=\mu \: \mbox{grad}\:p\:, \quad \mbox{div}\: \bv=0 \label{eq:stokes}\:,
\ee
respectively, accompanied by the boundary condition at the swimmer surface  $\tau=\tau_a$
\be
\bv=u(\zeta)\:\bbe_\zeta \:. \label{eq:bc}
\ee
Since the flow is axisymmetric we introduce the scalar stream-function $\Psi$ (unique up to an additive constant) that satisfies the continuity equation %in \ref{eq:stokes}
\be
\bv=v_\tau \bbe_\tau+v_\zeta \bbe_\zeta= \mbox{curl} \left(\frac{1}{\widehat{H}_\varphi}\: \Psi\: \bbe_{\varphi}\right) \label{eq:defv}\:.
\ee
The velocity components are readily obtained\ from \eqref{eq:defv} as
\[
v_\tau=\frac{1}{\widehat{H}_\zeta\widehat{H}_\varphi} \frac{\partial\Psi}{\partial\zeta}\:,
%=\frac{1}{c^2\{\tau^2-\zeta^2\}^\frac{1}{2}\{\tau^2-1\}^\frac{1}{2}}\:\frac{\partial\Psi}{\partial\zeta}\:, \label{eq:vtau}\\
v_\zeta=-\frac{1}{\widehat{H}_\zeta\widehat{H}_\varphi} \frac{\partial\Psi}{\partial\tau}\:,
%=\frac{-1}{c^2\{\tau^2-\zeta^2\}^\frac{1}{2}\{1-\zeta^2\}^\frac{1}{2}}\:\frac{\partial\Psi}{\partial\tau}\:\:.
%\label{eq:vzeta}
\]
where the symbols $\widehat{H}$ stand for the appropriate Lam\'e metric coefficients
$\widehat{H}_{\tau}=c(\tau^2-\zeta^2)^\frac{1}{2}(\tau^2-1)^{-\frac{1}{2}}$,   $\widehat{H}_{\zeta}=c(\tau^2-\zeta^2)^\frac{1}{2}(1-\zeta^2)^{-\frac{1}{2}}$,  $\widehat{H}_{\varphi}=c(\tau^2-1)^\frac{1}{2}(1-\zeta^2)^\frac{1}{2}$.
The vorticity field can be obtained from \eqref{eq:defv} as
\be
\te{\omega}=\mbox{curl}\:\bv=
%\mbox{curl curl}\left(\frac{1}{\widehat{H}_\varphi}\: \Psi\: \bbe_{\varphi}\right)
%=\omega_\varphi\bbe_\varphi
-\frac{1}{\widehat{H}_\varphi}E^2 \Psi \:\bbe_\varphi\:, \label{eq:vort}
\ee
where the operator $E^2$ is given by
\[
\mathrm{E}^2=\frac{1}{c^2(\tau^2-\zeta^2)}\left[(\tau^2-1)\frac{\partial^2}{\partial \tau^2}+(1-\zeta^2)\frac{\partial^2}{\partial \zeta^2} \right]. %\label{eq:E2}
\]
Following the standard procedure, the pressure is eliminated from the Stokes equation by applying the curl operator to both sides, with conjunction with \eqref{eq:vort} this yields the equation $\mathrm{E}^4\Psi=0$ for the stream-function. The boundary conditions \eqref{eq:bc} at the microswimmer surface $\tau=\tau_a$ in terms of the stream-function become
\be
\Psi=0\:, \qquad \partial_\tau \Psi=-c^2\{\tau_a^2-\zeta^2\}^\frac{1}{2}\{1-\zeta^2\}^\frac{1}{2}\: u(\zeta)\:, \label{eq:bc1}
\ee
and  the conditions at infinity ($\tau \rightarrow \infty$) are ${v_\tau}\sim -U\zeta$, ${v_\zeta}\sim -U(1-\zeta^2)^\frac{1}{2}$.

The solution for $\Psi$ that is regular on the axis and at infinity, and also even in $\zeta$ can be derived from a general \textit{semiseparable} solution \cite{DHP94,Zlat99},
\ba
\Psi=-2 c^2 U \:G_2(\tau) G_2(\zeta)+\sum\limits_{m=2,4,\ldots}^{\infty} \left\{A_m H_m(\tau) G_m(\zeta)+ \right.\nonumber \\
\left.C_m\:\Omega_m(\tau,\zeta)\right\},\label{eq:psi}
\ea
%where $\Omega_m(\tau,\zeta)$ is one the of the four eigenfunctions of $\mathrm{E}^4\Psi=0$
where $\Omega_m(\tau,\zeta)$ is a solution of $\mathrm{E}^4\Psi=0$ composed from spheroidal harmonics that decay at infinity,
%\[
%\Omega_2=\beta^*_2\left[H_2(\tau)G_4(\zeta)+H_4(\tau)G_2(\zeta)\right]+ \frac{1}{6}G_1(\tau)G_2(\zeta),
%\]
%\ba
%\Omega_m =-\alpha^*_m\left[H_{m-2}(\tau)G_m(\zeta)+H_m(\tau)G_{m-2}(\zeta)\right] && \nonumber \\
%+\beta^*_m\left[H_{m+2}(\tau)G_m(\zeta)+H_m(\tau)G_{m+2}(\zeta)\right],&& m \ge 4,  \nonumber
%\ea
%and the coefficients $\alpha^*_m$ and $\beta^*_m$ given by
%\[
%\alpha_m^*=\frac{(m-3)(m-2)}{2(2m-1)(2m-3)^2}\:,\qquad \beta_m^*=\frac{(m+1)(m+2)}{2(2m-1)(2m+1)^2}\:.
%\]
and $G_m$ and $H_m$ are the Gegenbauer functions of the first and the second kind, respectively.
%Note that \eqref{eq:psi} satisfies the condition at infinity, \eqref{eq:bcinf}.
The coefficients $A_m$ in \eqref{eq:psi} can be expressed in terms of $C_m$ and $U$ via the use of the boundary condition $\Psi=0$ at $\tau=\tau_a$.
Substituting \eqref{eq:psi} into \eqref{eq:bc1} we arrive after some algebra
%\[G^\prime_n(\tau)=-P_{n-1}(\tau)\:,\qquad H_n^\prime(\tau)=-Q_{n-1}(\tau)\:,
%\]
at the tridiagonal infinite system of equations for $U$ and the coefficients $C_m\:$,
\be
%& 2\:c^2 \:\lambda\:U + \gamma\:C_2 + {\cal E}_2^+\:C_4 = b_2, & \label{eq:m2} \\
{\cal E}_m^{(-)}C_{m-2}+{\cal E}_m^{(0)} C_m+{\cal E}_m^{(+)}\:C_{m+2}=b_m,\quad m\ge 2 \label{eq:mge4}
\ee
Here $C_0=-c^2U$, $\:{\cal E}^{(0,\pm)}_m$ are known functions of $\tau_a$, and
\be
b_m=\frac{c^2}{2}m(m-1)(2m-1)\int\limits_{-1}^{+1} \left\{\frac{\tau_a^2-\zeta^2}{1-\zeta^2}\right\}^{\frac{1}{2}} u(\zeta) G_m(\zeta) \:\rd \zeta. \label{eq:bm}
\ee
%and
%\ba
%&& \lambda= \left(P_1-\frac{G_2}{H_2}\:Q_1\right)\:,\nonumber \\
%&& \gamma = \beta^*_2\left(Q_3-\frac{H_4}{H_2}\:Q_1\right)+\frac{1}{6} \left(1-\frac{G_1}{H_2}\:Q_1\right)\:,\nonumber \\
%&& \delta = -\alpha^*_4\left(Q_3-\frac{H_4}{H_2}\:Q_1\right)\:, \nonumber \\
%&& {\cal E}^{(-)}=\beta^*_{m-2} \left(Q_{m-3}-\frac{H_{m-2}}{H_m}\:Q_{m-1}\right)\:, \nonumber \\
%&& {\cal E}^{(0)}=-\alpha^*_m \left(Q_{m-3}-\frac{H_{m-2}}{H_m}\:Q_{m-1} \right)+ \beta^*_m \left(Q_{m+1}-\frac{H_{m+2}}{H_m}\:Q_{m-1}\right)\:, %\nonumber \\
%&& {\cal E}^{(+)}=-\alpha^*_{m+2} \left(Q_{m+1}-\frac{H_{m+2}}{H_m}\:Q_{m-1}\right)\:, \nonumber
%\ea
%with $P_n,\: Q_n,\: G_n$ and  $H_n$ correspond to the values of the appropriate functions at $\tau=\tau_a$ and
%$\alpha^*_m=\frac{\alpha_m}{2(2m-3)}$, $\beta^*_m=\frac{\beta_m}{2(2m+1)}$.
Regularity of $\Psi$ implies that the admissible  solution of \eqref{eq:mge4} should satisfy
$C_mH_m(\tau_a)\rightarrow0$ as $m\rightarrow\infty$
while the exponentially growing solution with $C_m\sim \mathcal{O}((\tau_a+\sqrt{\tau_a^2-1})^{2m})$ should be discarded.

The viscous drag force exerted on the prolate spheroid (in the $x_3$-direction) is solely determined by the $C_2$-term in \eqref{eq:psi} corresponding to a monopole (Stokeslet) velocity term decaying like $1/r$ far from the particle, $F=-(4\pi\mu/c)\:C_2$ \cite{HB64}. Either $F$ or $U$ can be specified in addition to the surface velocity, $u(\zeta)$. In the swimming problem $F=C_2=0$, and $u(\zeta)$ determines the propulsion velocity $U$.

The problem of the ``frozen" spheroid (\emph{i.e.} $u(\zeta)=0$) in the uniform ambient flow $-U\bbe_3$ corresponds to  substituting $b_m=0,\: m \ge 2$ in \eqref{eq:mge4}.
%The solution to the problem of the flow around a ``frozen" spheroid (\emph{i.e.} $u(\zeta)=0$) in the uniform ambient flow $-U\bbe_3$ can be readily obtained by substituting $b_m=0,\: m \ge 2$ in \eqref{eq:mge4}.
%%%%%%%%%%%%%%%%%%%%%%%%%%%%%%%5
%In this case an exact solution can be found as \eqref{eq:mge4} are satisfied if the following recursive relation holds
%\[
%\alpha^*_m\:C_m-\beta^*_{m-2}\:C_{m-2}=0\:,\qquad m=4,\:6,\:\ldots\:.
%\]
%%%%%%%%%%%%
In this case the equations for $C_m$ can readily be solved yielding
%In this case it can be shown \cite{Zlat99} that $A_2$ and $C_2$ are the only terms that survive in \eqref{eq:psi} yielding
%substituting (\eqref{eq:CmCm-2}) into (\eqref{eq:psi}) and (\eqref{eq:m2}) we arrive at the
the well-known result for $\Psi$ and the drag force \cite{HB64},
\be
F=\frac{8 \pi c \mu U }{(1+\tau_a^2)\coth^{-1}{\tau_a}-\tau_a}. \label{eq:force}
\ee
%\be
%\Psi(\tau,\zeta)=\left\{2c^2\:U G_2(\tau)+A_2 H_2(\tau)+C_2 G_1(\tau) \right\}G_2(\zeta).\label{eq:psi0}
%\ee
%where the value of the coefficient $C_2$ and $A_2$ are determined from \eqref{eq:m2} and \eqref{eq:A2}.

\subsection{Propulsion velocity} 
\begin{figure}[t]
\begin{center}
\includegraphics[width=2.8in]{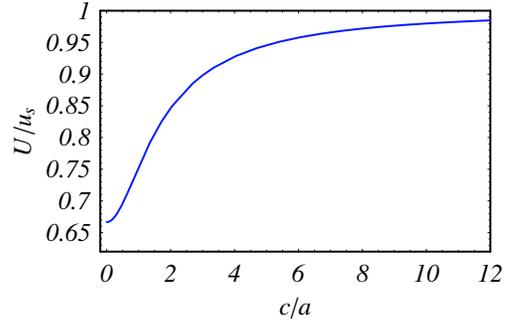}
\caption{The propulsion velocity of the `cigar-shaped' microswimmer vs. the elongation. \label{fig:Fig1}}
\end{center}
\end{figure}

In order to determine the velocity of propulsion of the microswimmer freely suspended in the viscous fluid, one must solve Eqs. \eqref{eq:mge4} with $C_2=0$ as the particle is force (and torque) free.
%An arbitrary surface velocity $u(\zeta)$ that meets all the above requirements for regularity and evenness in $\zeta$, can be expanded in the Gegenbauer polynomials $G_m(\zeta)$ according to \eqref{eq:bm} and the resulting tridiagonal linear system of equations must be solved for $U$ and $C_m$.
Let us consider the following velocity distribution at the boundary
\be
u(\zeta)=-2\tau_a u_s\:\left(\tau_a^2-\zeta^2\right)^{-\frac{1}{2}}\left(1-\zeta^2\right)^{-\frac{1}{2}}\:G_2(\zeta)\:,\label{eq:us}
\ee
where $u_s$ is a typical velocity of surface treadmilling and $G_2(\zeta)={1\over2}(1-\zeta^2)$. One may verify that for a sphere ($c/a\rightarrow 0$) $u(\zeta)=u_s\sin{\theta}$, while for an elongated swimmer $u(\zeta) \simeq u_s$ almost everywhere except the near vicinity of the poles $\zeta=\pm1$. More generally it can be readily shown that the solution satisfying \eqref{eq:bc1}, \eqref{eq:us} is given by \eqref{eq:psi} with $\forall m\:,C_m=0;m \ge 4,\: A_m=0$ and
\ba
&& A_2=-2 c^2 u_s \tau_a \:(-1+\tau_a^2), \nonumber \\
&& U=u_s\left\{\tau_a^2-\tau_a\:(-1+\tau_a^2)\coth^{-1}\tau_a\right\}. \label{eq:U}
\ea
Note that $U$ can be related to the surface motion\footnote{In the laboratory frame the velocity at the surface is a superposition of the translational velocity $\bU$ and purely tangential motions $\bu$} via the use of the Lorentz reciprocal theorem \cite{KK91},
\be
\widehat{\bF}\cdot\bU=-\int_S (\widehat{\te{\sigma}}\cdot\bn)\cdot \bu\:\rd S\:,\label{eq:recip}
\ee
where $(\widehat{\te{u}},\: \widehat{\te{\sigma}})$ is the velocity and stress field corresponding to translation of the same shaped object when acted upon by an external force $\widehat{\bF}$. For purely tangential surface motion considered in this work we have $(\widehat{\te{\sigma}}\cdot\bn)\cdot \bu=\widehat\sigma_{\tau\zeta}\:u(\zeta)$, where
\[
\widehat\sigma_{\tau\zeta}=\frac{\widehat{H}_\tau}{\widehat{H}_\zeta}\frac{\partial}{\partial\zeta}\left(\frac{\hat{v}_\tau}
{\widehat{H}_\tau}\right)+\frac{\widehat{H}_\zeta}{\widehat{H}_\tau}\frac{\partial}{\partial\tau}\left(\frac{\hat{v}_\zeta}
{\widehat{H}_\zeta}\right).
\]
We calculate the local tangential stress component $\widehat\sigma_{\tau\zeta}$ from the solution corresponding to streaming past a rigid prolate spheroid, while $\widehat{F}$ is given by \eqref{eq:force}.  Substitution into \eqref{eq:recip} with $\rd S={\widehat{H}_\varphi\widehat{H}_\zeta}\: \rd \varphi\: \rd \zeta$ yields after some algebra
\be
U=-\frac{\tau_a}{2}\:\int\limits_{-1}^{+1}\left(\frac{1-\zeta^2}{\tau_a^2-\zeta^2}\right)^{\frac{1}{2}}u(\zeta)\: \rd\zeta,\label{eq:recip1}
\ee
which holds for an arbitrary tangential boundary velocity $u(\zeta)$. In the special case of $u(\zeta)$ given by \eqref{eq:us} it can be readily demonstrated that integration yields the propulsion speed \eqref{eq:U}. Also, \eqref{eq:recip1} actually solves the infinite tridiagonal system \eqref{eq:mge4}, since knowing $U,F$ (i.e. $C_0,C_2$) one can iteratively obtain all the other $C_m$'s by direct substitution.
The scaled swimming speed of the microswimmer is depicted in Figure \ref{fig:Fig1} as a function of the scaled elongation.
The values of the propulsion velocity corresponding to a spherical swimmer ($c=0$) and a slender
swimmer ($c\gg a$) can be determined via  \eqref{eq:recip} without invoking special spheroidal coordinates.
%\be
%\widehat{\bF}\cdot\bU=-\int_S (\widehat{\te{\sigma}}\cdot\bn)\cdot \bu\:\rd S\:,\label{eq:recip}
%\ee
%where $\widehat{\te{\sigma}}$ is the stress field corresponding to the translation of the same shaped object when acted upon by an external force $\widehat{\bF}$.
For a sphere, the local traction force $\widehat{\te{\sigma}}\cdot\bn=-\frac{3\mu}{2a}\hat{\bU}$ and $\hat{\bF}=-6\pi\mu a\hat{\bU}$ and thus
the self-propulsion velocity can be found as \cite{SS96}
\[
U=-\frac{1}{4\pi a^2}\int_S \bu\cdot\bbe\: \rd S, % \label{eq:Usph}
\]
where $\bbe$ is the unit vector in the direction of locomotion. Substituting $\bu=u_s \sin\theta\: \bbe_\theta$ and $\bbe=\bbe_r\:\cos{\theta}-\bbe_\theta\:\sin\theta$, where $\theta$ is the spherical angle measured with respect to $\bbe$, we arrive at $U=\frac{1}{2} \int_{0}^{\pi} u_s \sin^3\theta \:\rd \theta=\frac{2}{3}\:u_s$ in agreement with the result shown in Figure \ref{fig:Fig1}.

The drag force exerted on the rod-like microswimmer upon translation along its major axis with velocity $\hat{\bU}^{||}$ is given by $\hat{\bF}\approx %4\pi\mu b\hat{\bU}^{||}/\log{(b/a)}=
-4\pi\mu a \hat{\bU}^{||}/(\varepsilon \log{1/\varepsilon})$ and the local friction force is $\widehat{\te{\sigma}}\cdot\bn \approx -\mu\hat{\bU}^{||}/(a\log{1/\varepsilon})$, where $\varepsilon=a/b=[1+(c/a)^2]^{-1/2}\ll 1$ is the aspect ratio. Thus, from \eqref{eq:recip} follows
\[
U \approx -\frac{1}{4\pi a b}\int_S \bu\cdot\bbe\:\rd S. %\label{eq:Urod}
\]
For the `needle-shaped' microswimmer the surface velocity $\bu=u(\zeta)\bbe_{\zeta}\simeq -u_s\: \bbe$ over almost the whole surface, it follows that $U \simeq u_s$. As seen in Figure \ref{fig:Fig1} the propulsion velocity $U/u_s\rightarrow 1$  as $c/a$ grows and equals to $0.95$ already at $c \simeq 5.3a$. As intuitively expected, the micro-swimmer is self-propelled forward with the velocity of the surface treadmilling
%(in the coordinate frame co-moving with the micro-swimmer)
, while the boundary velocity in the laboratory frame is  (almost) zero.
%The fluid around the thin rod-like swimmer remains undistorted except, perhaps, for the near vicinity of the poles, \textit{i.e.} the swimmer is propelled forward throughout the almost quiescent fluid.

\subsection{Swimming efficiency} 
\begin{figure}[t]
\begin{center}
\includegraphics[width=2.8in]{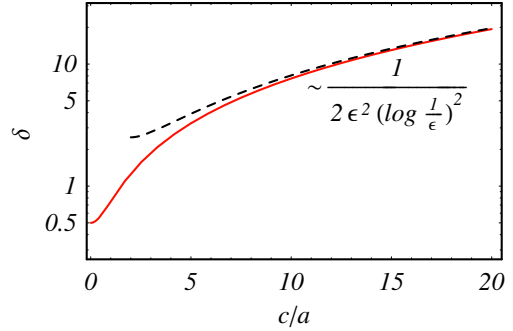}
\caption{The swimming efficiency of the spheroidal microswimmer, $\delta$, vs. the scaled elongation $c/a$ (linear-log plot). The solid line corresponds to the exact calculation, the dashed line is the asymptotic result.\label{fig:Fig2}}
\end{center}
\end{figure}

Since the fluid around the elongated microswimmer propelled by continuous surface treadmilling is almost quiescent, except for the near vicinity of the poles, it is natural to expect low viscous dissipation and high \emph{hydrodynamic} swimming efficiency. Several definitions of hydrodynamic efficiency have been proposed \cite{SS96,SW89,AGK04} here we follow the definition $\delta=\bF\te{\cdot}\:\bU/{\cal P}$, where ${\cal P}$ is the energy dissipated in swimming with velocity $\bU$, and the expression in the numerator is the work expanded by dragging the ``frozen" swimmer at velocity $\bU$ upon action of an external force $\bF$ \cite{SS96}. $\delta$ is dimensionless and compares the self-propulsion with dragging (some authors use  the reciprocal efficiency $1/\delta$). The higher $\delta$ the more efficient the swimmer is.
For an axisymmetric swimmer propelled along the symmetry axis, $\bF\te{\cdot}\:\bU={\cal R} U^2$, where the scalar ${\cal R}$
%\mbox{\boldmath$\cal R$}_{FU}\te{\cdot}\:\bU=
is the appropriate hydrodynamic resistance.
The work done by an arbitrary shaped swimmer and dissipated by viscosity in the fluid is given by
\[
{\cal P}=-\int_S (\te{\sigma}\cdot\bn)\cdot \bv\: \rd S=2\mu \int_V \bE \te{:} \bE \: \rd V\:, %\label{eq:dissp}
\]
where $\bE$ is the rate-of-strain tensor, $V$ is the fluid volume surrounding the swimmer and $S$ its surface.
Expressing the product $\bE\te{:}\bE$ as $\sum \omega_i \omega_i+2 (\partial_i v_j) (\partial_j v_i)$ allows re-writing  ${\cal P}$ for microswimmers self-propelled by purely tangential motions $\bu$ as\cite{SS96}
\be
{\cal P}=\mu\int_V\te{\omega}^2 \rd V+2\mu \int_S\:\bu^2 \kappa_s\:\rd S\:, \label{eq:dissp2}
\ee
where $\kappa_s=-({\partial \bs}/{\partial s})\te{\cdot}\:\bn$ is the curvature measured along the path of the surface flow, expressible in terms of the unit tangential and normal vectors, $\bs$ and $\bn$, respectively.
%This expression for ${\cal P}$ can be applied to verify that for the rod-like treadmiler with rounded ends  $\delta\sim(\varepsilon \log{1/\varepsilon})^{-1}$ as was derived from simple scaling arguments in the introduction. Over the straight portion of the swimmer the surface velocity is $\bu \simeq -u_s\:\bbe_3$ and from \eqref{eq:Urod} the swimming speed, $U \approx u_s$. Neglecting friction associated with the rounded ends (denoted$ \pm$) the work in dragging can be estimated as ${\cal R}U^2 \simeq \frac{4 \pi \mu b U^2}{\log{(2/\varepsilon)}} \sim \frac{\mu a U^2}{\varepsilon \log{1/\varepsilon}}$. Over the straight portion of the treadmiler $\kappa_s=0$ and for the ends $\kappa_s=1/a$, and from \eqref{eq:dissp2} follows
%\be
%{\cal P} \ge 2\mu\:\int_{S} \kappa_s \bu^2\: \rd S \ge \frac{2\mu}{a} \int_{S_\pm} \bu^2\: \rd S =\mathcal{O}(\mu a U^2)\:. \label{eq:dissp4}
%\ee
%Thus, the hydrodynamic efficiency in this case is $\delta={\cal R}U^2/\mathcal{P}\sim (\varepsilon \log{1/\varepsilon})^{-1}$.
Let us now estimate $\delta$ of the spheroidal treadmiler described in the previous subsection. For a prolate spheroid
%the unit tangential and normal vectors
$\bs=\bbe_\zeta$, $\bn=\bbe_\tau$, respectively, and $\kappa_s$ can be calculated as
\[
\kappa_s=\left.\frac{1}{{\widehat{H}_\tau\widehat{H}_\zeta}} \frac{\partial \widehat{H}_\zeta}{\partial \tau}\right|_{\tau=\tau_a}=\frac{\tau_a \sqrt{\tau_a^2-1}}{c (\tau_a^2-\zeta^2)^{3/2}}\:,
\]
Since the solution \eqref{eq:U} corresponds to irrotational flow\footnote{Since $\mathrm{E}^2(H_m G_m)=0$ the votricity $\te{\omega}$ is determined by the $C_m,\:m \ge 4$ terms in \eqref{eq:psi}}, \emph{i.e.} $\te{\omega}=0$  the volume integral in \eqref{eq:dissp2} drops out. Substituting the expression for the surface velocity \eqref{eq:us} and $\kappa_s$ into the surface integral in \eqref{eq:dissp2} we find
\be
{\cal P}=
%4\pi\mu \int_{-1}^{+1} u^2(\zeta)\: \kappa_s\: \widehat{H}_\varphi\:\widehat{H}_\zeta\:\rd \zeta=
4\pi \mu c u_s^2 (\tau_a^2-1)\:\left\{(1+\tau_a^2)\coth^{-1}\tau_a-\tau_a\right\}. \label{eq:dissp3}
\ee
Collecting the expressions for the drag force \eqref{eq:force}, velocity of self-propulsion \eqref{eq:U} and the dissipation \eqref{eq:dissp3} one can compute the swimming efficiency,
\be
\delta=\frac{{\cal R} U^2}{\cal P} =\frac{2\left\{\tau_a^2-\tau_a(\tau_a^2-1)\coth^{-1}\tau_a\right\}^2}{(\tau_a^2-1)\:\left\{(1+\tau_a^2)\coth^{-1}\tau_a-
\tau_a\right\}^2}\:. \label{eq:delta}
\ee
$\delta$ is plotted as a function of the elongation $c/a$ in Figure \ref{fig:Fig2}.
Evidently, $\delta$ grows unbounded as $c/a \rightarrow \infty$ does, corresponding, in the limit to a frictionless swimmer. For the spherical treadmiler $\delta$ can be calculated from \eqref{eq:dissp2} with $\bu=u_s\:\sin\theta\:\bbe_\theta$, $u_s=\frac{3}{2}U$ and $\kappa_s=1/a$,
$
{\cal P}=2\mu\:\int_S \frac{9}{4} \:U^2\: \sin^2{\theta}\: \frac{1}{a}\: \rd S =12\:\pi\mu a U^2$.
Dividing $6\pi\mu a U^2$ by ${\cal P}$ we find $\delta=\frac{1}{2}$ in agreement with \eqref{eq:delta} (see Figure \ref{fig:Fig2}) and the theoretical bound (\textit{i.e.} $\delta \le \frac{3}{4}$) in \cite{SS96} .

For the slender swimmer the asymptotic behavior of $\delta$ can be estimated from \eqref{eq:delta} by expanding $\delta$ in a series around $\tau_a=1$ and using $\tau_a=1/\sqrt{1-\varepsilon^2}\sim 1+\frac{\varepsilon^2}{2}$ where $\varepsilon=\frac{a}{b} \ll 1$,
\be
\delta\simeq \frac{1}{2} \frac{1}{(\varepsilon\log{\varepsilon})^2}.
%+{\cal O}\left(\frac{1}{(\log{1/\varepsilon})^2}\right).
\label{eq:delta2}
\ee
%where as before $E=\left(\log{\frac{2}{\varepsilon}}-\frac{1}{2}\right)^{-1}\ll 1$.
This result is shown in Figure \ref{fig:Fig2} as a dashed line. For comparison, the efficiency of spherical squirmers self-propelled by propagating surface waves along their surface (the mathematical model of cianobacteria \cite{ESBM96}) has the upper bound $\delta \le \frac{3}{4}$, while numerically calculated values of $\delta$ do much worse than dragging and the corresponding swimming efficiency is usually less than 2\% \cite{SS96}. Swimming by surface treadmilling is remarkably more efficient than the rotating helical flagellum \cite{percel97}, beating flexible filament \cite{WG98}, the Percell's ``three-link swimmer" \cite{BKS03} or locomotion by virtue of shape strokes \cite{SW89, AGK04}. The surface treadmilling is probably superior to any inertialess swimming techniques proposed so far.

Also, the swimming efficiency of the ellipsoidal treadmiler is superior by a factor of $(\varepsilon\log{1/\varepsilon})^{-1}$ over the estimate of $\delta$ corresponding to the rod-like treadmiler with rounded ends derived from purely scaling arguments in the introduction. Therefore, the geometry (via $\kappa_s$) plays an important role in minimizing the dissipation in surface treadmilling, which is rather surprising since the drag force on slender nonmotile object does not depend on its shape to the first approximation.

\subsection{Optimal swimming}
We can set an upper bound on $\delta$ for a spheroidal microswimmer in terms of surface integrals of an \emph{arbitrary} velocity $u(\zeta)$ analogously to \cite{SS96}.
The power dissipated in self-propulsion is bounded from below according to \eqref{eq:dissp2} by
\[
{\cal P} \ge 2\mu \int \bu^2 \kappa_s \rd S= 4\pi\mu c\: \tau_a \left(\tau_a^2-1 \right)\int\limits_{-1}^{+1}\frac{u^2(\zeta)}{\tau_a^2-\zeta^2}\:\rd \zeta,
\]
where we used the previously derived result for $\kappa_s$.
The power expanded in dragging at the same speed is found from \eqref{eq:force} and \eqref{eq:recip1} as
\be
{\cal R}U^2=\frac{2\pi c \mu \tau_a^2}{(1+\tau_a^2)\coth^{-1}\tau_a-\tau_a}\left[ \int\limits_{-1}^{+1}\left\{\frac{1-\zeta^2}{\tau_a^2-\zeta^2}\right\}^{\frac{1}{2}}u(\zeta)\rd\zeta\right]^2. \nonumber
\ee
Combining the last two results we obtain an upper bound on $\delta$ as
\[
\delta \le \frac{\tau_a}{(\tau_a^2-1)\left((1+\tau_a^2)\coth^{-1}\tau_a-\tau_a\right)}\textstyle\left\{\frac{\left[ \int\limits_{-1}^{+1}\left(\frac{1-\zeta^2}{\tau_a^2-\zeta^2}\right)^{\frac{1}{2}}u\: \rd\zeta\right]^2}{2 \int\limits_{-1}^{+1}\frac{u^2}{\tau_a^2-\zeta^2}\:\rd \zeta}\right\}
\]
The term in the figure brackets can be shown to be bounded from above by $2/3$ while its maximum is obtained for $u(\zeta)=u_s\sqrt{1-\zeta^2}\sqrt{\tau_a^2-\zeta^2}$ corresponding to the $2$-term boundary velocity expansion \eqref{eq:mge4} with $b_2,\:b_4 \ne 0\:,b_m=0, m \ge 6$ (and, thus, representing a rotational flow). Thus, for an elongated swimmer ($\tau_a \rightarrow 1$) we arrive at
\[
\delta\leq\frac{1}{3\varepsilon^2 \log{1/\varepsilon}}
\]
which does better than \eqref{eq:delta2} by a factor of $(\log{1/\varepsilon})^{-1}$ and also superior over the scaling estimate for the rod-like swimmer by a factor of $\mathcal{O}(1/\varepsilon)$. Note that the  asymptotic behavior $\delta\sim\frac{1}{\varepsilon^2\log{1/\varepsilon}}$ was derived from simple scaling arguments in the introduction.

It can be demonstrated that the surface velocity \eqref{eq:us} is not optimal, \emph{i.e.} it does not minimize ${\mathcal P}$ for a prescribed propulsion speed $U$. To see this consider the slightly perturbed boundary velocity
\be
u(\zeta)=\frac{-2\tau_a}{\left(\tau_a^2-\zeta^2\right)^{\frac{1}{2}}\left(1-\zeta^2\right)^{\frac{1}{2}}}\left\{u_2 G_2(\zeta)+
u_4 G_4(\zeta)\right\},\label{eq:us2}
\ee
such that $|u_4| \ll |u_2|$ and $u_2\sim u_s$. The solution of the linear problem \eqref{eq:mge4} yields $|\te{\omega}|={\cal O}(u_4)$  and therefore, the volume integral in \eqref{eq:dissp2} is
$\mu \int_V \te{\omega}^2  \rd V={\cal O}(u_4^2)$. The surface integral in \eqref{eq:dissp2} can be calculated as
\[
2\mu \int_S \bu^2 \kappa_s \rd S=u_2^2{\cal I}_2(\tau_a) + u_2 u_4 {\cal I}_4(\tau_a),
\]
where ${\cal I}_2(\tau_a)$ is given by \eqref{eq:dissp3} and ${\cal I}_4$ is some other function of $\tau_a$.
%\[
%{\cal I}_4=2\pi \mu c(\tau_a^2-1)\left(-\tau_a+15\tau_a^3+(1+6\tau_a^2-15\tau_a^4)\coth^{-1}\tau_a\right)\:.
%\]
The velocity of propulsion can be found in the close form as $U=u_2 {\cal F}_2(\tau_a)+u_4 {\cal F}_4(\tau_a)$ from \eqref{eq:recip1}, where ${\cal F}_2$ is equal to the expression in the figure brackets in \eqref{eq:U} and ${\cal F}_4$ is some other function of $\tau_a$.
%\[
%{\cal F}_4=-\frac{1}{12}\:\tau_a\left(13\tau_a-15\tau_a^3+3(1-6\tau_a^2+5\tau_a^4) \coth^{-1}\tau_a\right)\:.
%\]
As the  propulsion velocity to be fixed, we require $U=u_s {\cal F}_2$. This yields dissipation
%$u_2=u_s-\epsilon\:({\cal F}_4/{\cal F}_2)$. Substituting this into \eqref{eq:dissp3} yields
$$
{\cal P}=u_s^2 {\cal I}_2 +  u_4 u_s \left({{\cal I}_4-2 \frac{{\cal F}_4}{{\cal F}_2}\: {\cal I}_2} \right)+{\mathcal O}(u_4^2).
$$
where the function in the brackets
%\[
%\frac{20 \pi \mu c\:(\tau_a^2-1)\:\left[\tau_a+3\:\tau_a^3+\coth^{-1}\tau_a \left(-1+5\tau_a^2-6\tau_a^4+3\tau_a %(\tau_a^2-1)^2\coth^{-1}\tau_a\right) \right]}{\left(-3\tau_a+3(\tau_a^2-1)\coth^{-1}\tau_a\right)}
%\]
can be shown to be positive and bounded for all $\tau_a>1$, and vanishes only at $\tau_a \rightarrow 1$. Therefore, one can always choose some $u_4<0$ such that $P<{\cal I}_2 u_s^2$ leading to reduction in the dissipation in \eqref{eq:dissp3}. The above perturbation analysis shows that, quite surprisingly, vorticity production could bring a reduction in viscous dissipation, leading to more efficient swimming.

To address the question of optimal swimming we consider an arbitrary boundary velocity via the  expansion, that meets all the above requirements for regularity and evenness in $\zeta$,
\be
u(\zeta)=\frac{-2\tau_a}{\left(\tau_a^2-\zeta^2\right)^{\frac{1}{2}}\left(1-\zeta^2\right)^{\frac{1}{2}}}\:\sum_{m=2,4,\ldots}^L u_m G_m(\zeta).\label{eq:us1}
\ee
where it follows from \eqref{eq:bm} that $u_m=-b_m/(2 c^2 \tau_a)$. To find a set of Fourier coefficients $u_m,\:m=2,\: 4,\:6, \ldots\:L$ corresponding to the optimal swimming, one should minimize the dissipation integral, ${\cal P}$, while keeping the propulsion speed $U$ fixed. The dissipation integral ${\cal P}=-\int_S \sigma_{\tau\zeta}\: u\: \rd S$ being bilinear in $u_i$, can be expressed as $\mathcal{P}={1\over2}\sum u_i \:{\mathcal P}_{ij} \:u_j$.
%(the summation agreement applies hereafter).
Note however that the tangential stress $\sigma_{\tau\zeta}$ at the surface of the microswimmer requires the knowledge of the velocity gradient at the surface (rather than velocity along).
Alternatively, since the optimal velocity field is rotational, calculation of
${\cal P}$ from \eqref{eq:dissp2} requires the knowledge of vorticity everywhere.
%The vorticity $\te{\omega}$ can be found from (\eqref{eq:vort}) is a linear form of acting on $\{C_i\}$ that can be determined via the solution of the linear system of equations (\eqref{eq:m2}--\eqref{eq:mge4}) with $b_m=2 c^2 \tau_a\:u_m$ from (\eqref{eq:bm}) and (\eqref{eq:us1}).

The propulsion velocity given by \eqref{eq:recip1} is linear in $u_i$, \emph{i.e.} $U=\sum_j {\cal F}_j\: u_j$. The optimal set of coefficients $u_i$ is to be determined from $\frac{\partial}{\partial u_i} (\mathcal{P}-\lambda U)=0$, or just from $\sum_j {\mathcal P}_{ij} \:u_j=\lambda\:{\mathcal F}_i$, where $\lambda$ is a Lagrange multiplier.
%We can further eliminate $\lambda$ by proper re-scaling of $u_i$ and then solve the resulting linear system of equations for $u_i$. Then $\lambda$ to be determined from the condition $U={\mathcal F}_2 u_s$, resulting in $\lambda={\mathcal F}_2/\sum {\mathcal F}_m (u_m/u_s)$.
%
\begin{figure}[t]
\centering{\includegraphics[width=2.8in]{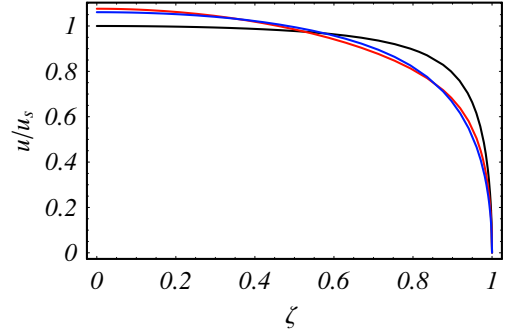}}
\caption{The optimal boundary velocity $u$ vs. the spheroidal coordinate $\zeta$ for elongation $c=2.5\:a$ at various truncation levels : $L=4$ (red) and $L=10$ (blue). The black line correspond to the one-term boundary velocity \eqref{eq:us}\label{fig:Fig3}}
\end{figure}
We found the closed form optimal solution for the two-term boundary velocity \eqref{eq:us1}, while for $L>4$ closed form expressions are cumbersome, and numerical solutions were derived instead.
Analogously to the theory for a 2-D swimmer (see the next section), where the explicit optimal solution was shown to acquire an infinite number of harmonics in the expansion for the boundary velocity, increasing the truncation level $L$ in \eqref{eq:us1} will further improve the efficiency of swimming, though the enhancement appears to be minor. To illustrate this, we calculate the optimal solution upon varying $L$. The optimal boundary velocity upon varying $L$ is depicted in Figure \ref{fig:Fig3} for the elongation of $c=2.5a$ and compared with the one-term expression \eqref{eq:us}. The scaled dissipation integral, ${\cal P}/\mu a u_s^2$, is plotted vs. $c/a$ upon varying $L$ in Figure \ref{fig:Fig4}. It can be readily seen that the convergence with respect to $L$ is rather fast; the deviation between the results corresponding to $L=8$ and $10$ is less than $1$\% for all $c/a$ and it vanishes at both limits $c=0$ and $c/a\rightarrow \infty$. Thus, the `intuitive' one-term boundary velocity \eqref{eq:us}, that yields $\delta \sim (\varepsilon \log{\varepsilon})^{-2}$ (see Figure \ref{fig:Fig2}) is nearly optimal for a wide range of elongations and likely so for all elongations.
%quite close to the optimal solution for all elongations. Also, for the cigar-shaped treadmiler the optimal propulsion efficiency scales as $(\varepsilon \log{1/\varepsilon})^{-2}$ and thus \eqref{eq:us} is an \textit{asymptotically} optimal surface velocity. This result is expected, since $u(\zeta)\simeq u_s$ a.e. over the surface of the microswimmer and the exact behavior of $u$ near the poles should not matter for a very long and thin microswimmer.
%
\begin{figure}[t]
\centering{\includegraphics[width=2.8in]{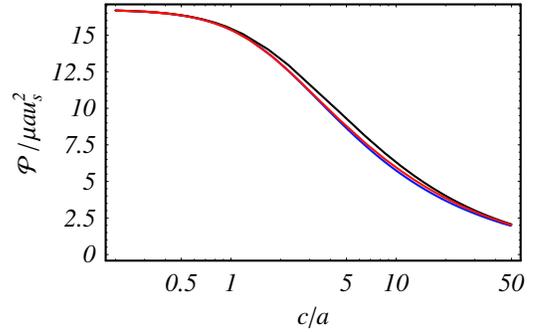}}
\caption{The dissipation integral vs. the scaled elongation corresponding to optimal swimming at various truncation levels (linear-log plot): $L=4$ (red) and $L=10$ (blue). The black line refers to \eqref{eq:dissp3}. \label{fig:Fig4}}
\end{figure}

\subsection{2-D microswimmer}
The two dimensional Stokes equations %$\mu\nabla^2\vec{v}=\vec{\nabla}P,\; div(\vec{v})=0$ \eqref{eq:stokes}
is conveniently handled by employing complex variables
%can be solved in general form employing complex variables
\cite{SW89,AGK04,Rich68,JM92}. This allows explicit solution of the optimization problem for the elliptical treadmiler.

Denoting $v=v_x+\ri v_y$ and $\partial={1\over2}(\partial_x-\ri\partial_y)$, the Stokes equations become $2\mu\partial\bar{\partial}v=\bar{\partial}p,\; \rRe\partial v=0.$ The most general solution to this (with $p$ real) is $v=g+\bar{f}-z\overline{g'},\; p=-4\mu \rRe(g')$ where $g,f$ are any pair of holomorphic functions \cite{Langlois64}. Solutions corresponding to multivalued $g,\:f$ are also legitimate (provided the resulting $v,\:p$ are single valued).  It can be shown (using\eqref{eq:dF} below) that the monodromy of $g$ (and of $\bar{f}$) around a closed curve gives the total force exerted by the fluid on the interior of the curve. In particular in swimming problems this force must vanish and $g,f$ are therefore always single valued.

The element of force $\rd F\equiv \rd F_x+\ri \rd F_y$ acting on a length element $\rd z=\rd x+\ri \rd y$ of the fluid can be expressed in terms of $v,P$ and hence in terms of $g,f$. Straightforward calculation shows that the relation is
\be
\rd F=\ri p \:\rd z+(2\ri\mu\bar{\partial}v)\:\rd\bar{z}=2\ri\mu\: \rd(v-2g). \label{eq:dF}
\ee
Note that here $(\rd x,\rd y)$ is tangent rather then the normal to the segment\footnote{The sign convention here is that $dF$ is the force exerted by the l.h.s of the (oriented) segment $dz$ on its r.h.s.}.

We consider a 2-D swimmer shaped as an ellipse of semi-axes $b,a=1\pm\alpha$ (with $0\le\alpha<1$) situated in the complex $z=x+\ri y$ plane. It is then convenient to define a new complex coordinate $\zeta$ by the relation $z=\zeta+\alpha/\zeta$. As $\zeta$ ranges over the region $|\zeta|>1$ the corresponding $z$
ranges over the area outside the swimmer. In particular the swimmer boundary corresponds to the unit circle $\zeta=\re^{\ri\theta}$. Note that if we consider $g,\:f$ as functions of $\zeta$ rather then $z$ then the general solution of the Stokes equations becomes
\be
v=g+\bar{f}-{\zeta+\alpha/\zeta\over 1-\alpha/\bar{\zeta}^2}\: \overline{g'}. \label{eq:v}
\ee

In the swimmer frame of reference, the boundary condition at infinity $v(\infty)=-U$ (where $U$ is the laboratory-frame swimming speed) implies Laurent expansions
\be
f=\sum_{n=1}^\infty a_n{\zeta}^{-n},\, g=-U+\sum_{n=1}^\infty b_n{\zeta}^{-n}, \label{eq:laurent}
\ee
where $U$ is arbitrarily appended to $g$.

The boundary condition on the swimmer surface is fulfilled by matching $v(\zeta)$ to a prescribed boundary motion $v|_{\zeta=\re^{\ri\theta}}=w(\theta)=\sum_{n=-\infty}^\infty w_n \re^{\ri n\theta}$.
It is useful to express $w(\theta)$ as $w=\overline{w_{+}(\zeta)}+w_{-}(\zeta),\zeta=e^{\i\theta}$ where $w_{-}=\sum_{n=0}^\infty w_{-n}{\zeta}^{-n},w_{+}=\sum_{n=1}^\infty w_{n}^*{\zeta}^{-n}$ are both analytic outside the unit circle.
Substituting \eqref{eq:laurent} into \eqref{eq:v} and matching on the unit circle we find
\be
g(\zeta)=w_{-}(\zeta),\,   f(\zeta)=w_{+}(\zeta)+{\zeta(1+\alpha\zeta^2)\over\zeta^2-\alpha}w_{-}'(\zeta). \nonumber
\ee
%\be
%g(\zeta)=\sum_{n=0}^\infty w_{-n}{\zeta}^{-n},\,   f(\zeta)=\sum_{n=1}^\infty  w_{n}^*{\zeta}^{-n}+{\zeta(1+\alpha\zeta^2)\over\zeta^2-\alpha}g'(\zeta). \nonumber
%\ee
In particular the swimming velocity is determined by the constant term in this expansion $U=-w_0$.
The corresponding dissipation is calculated using \eqref{eq:dF} as
\[
{\mathcal P}=-\rRe\oint\bar{v} \rd F=2\mu\: \rIm \oint\bar{w}\: \rd(2g-w)=4\pi\mu\:\sum_{n=-\infty}^\infty |n||w_n|^2.
%2\mu\: \rIm \oint(\overline{w_{-}}+w_{+})\: \rd(w_{-}-\overline{w_{+}})
\]

Let us next focus on the case of an ellipse swimming by surface treadmilling. The boundary velocity $w(\theta)$ being tangent to the swimmer boundary is expressible as $w={\rd z\over \rd\theta}u(\theta)=\ri(\zeta-\alpha/\zeta)u(\theta)$ for some real-valued function $u(\theta)$. Since we consider only swimmers symmetric with respect to the $x-$axis, we assume $u(\theta)$ to be an odd function allowing to write it as $u=\sum u_n\sin(n\theta)={1\over2\ri}\sum u_n(\zeta^n-\zeta^{-n})$.
In terms of this the swimming velocity turn into $U=-w_0={1\over2}(1+\alpha)u_1$ while the dissipation takes the form
$${\mathcal P}=2\pi\mu\sum n\left(  (1+\alpha^2)u_n^2-2\alpha u_{n-1}u_{n+1}\right).$$
Which may also be written as $\mathcal{P}= {1\over2}\sum \mathcal{P}_{ij}u_iu_j$ for a corresponding tridiagonal matrix $\mathcal{P}_{ij}$.

The optimal swimming technique for a given $\alpha$ is the one that minimizes the dissipation while keeping the swimming velocity
$U={1\over 2}(1+\alpha)u_1$ fixed. The minimizer is the solution of
${\partial\over\partial u_i}(\mathcal{P}-\lambda u_1)=0$ for $\forall i$ with $\lambda$ being a Lagrange multiplier, or just
\be
\sum_j \mathcal{P}_{ij}u_j=\lambda\delta_{i,1}. \label{eq:Pij}
\ee

It is readily seen that the coefficients $u_k$ with even $k$ are not relevant to the optimal swimming and should be set to zero to minimize viscous dissipation. (This is also clear from the fact that $u_{2k}$ correspond to flows which are antisymmetric with respect to the $y-$axis.)

Denoting $b_k\equiv u_{2k+1},\; k\geq 0$ and
%choosing a convenient normalization $\lambda=2\pi\mu(1+\alpha^2)$
writing $\lambda=2u_s\pi\mu(1+\alpha^2)$ with $u_s$ an arbitrary normalization constant having dimensions of velocity
we obtain from \eqref{eq:Pij} the  recursion relation
$$(2k+1)b_k-\xi (kb_{k-1}+(k+1)b_{k+1})=u_s\delta_{k,0},$$
where $\xi={2\alpha\over 1+\alpha^2}$. Multiplying by $x^k$ and summing over $k$ this transforms into a differential equation for the generating function $B(x)={1\over u_s}\sum_{k=0}^\infty b_kx^k$,
$$h(x)B'(x)+{1\over 2}h'(x)B(x)+1=0\:,$$
where we defined $h(x)%=\xi x^2-2x+\xi
\equiv\xi(x-\alpha)(x-{1\over\alpha})$. The general solution to this is $B(x)=-{1\over\sqrt{h(x)}}\int_C^x{\rd x'\over\sqrt{h(x')}}$ where $C$ is a constant of integration. Requiring the coefficients $b_k$ to decay for large $k$ implies that $B(x)$ must be analytic inside the unit disc and hence its potential singularity at $x=\alpha$ must be avoided. This determines the integration constant to be $C=\alpha$, so that we may write
\be
B(x)=-{1\over\sqrt{h(x)}}\int_{\alpha}^x{\rd x'\over\sqrt{h(x')}}={2\over\xi\sqrt{x_-x_+}}\log
\left( {\sqrt{x_-}+\sqrt{x_+}\over\sqrt{x_+-x_-}}\right), \nonumber
\ee
where $x_+={1\over\alpha}-x$ and $x_-=\alpha-x$.
The corresponding swimming velocity and the dissipation are, respectively,
\ba
U=\frac{(1+\alpha)u_1}{2}={u_s\over2}(1+\alpha) B(0)={u_s(1+\alpha)\over 2\xi} \log\left({1+\alpha\over 1-\alpha}\right),  \nonumber \\
\mathcal{P}={1\over2}\sum P_{ij}u_i u_j={1\over2}\lambda u_1=\pi\mu(1+\alpha^2)u_s^2B(0)=  \qquad \qquad \; \nonumber \\
{\pi\mu\over2\alpha}u_s^2(1+\alpha^2)^2\log{\left({1+\alpha\over 1-\alpha}\right)}.  \qquad \qquad \qquad \nonumber
\ea
Therefore, combining the last two results yields
\[
{\mathcal{P}}= { 2\alpha\over(1+\alpha)^2}\frac{4\pi\mu U^2}{\log\left({1+\alpha\over 1-\alpha}\right)}
\]
We recall that in 2-D the dragging problem admit no regular solution
within the Stokes approximation \footnote{ This is known as the Stokes paradox and can be resolved by noting that far from the object the quadratic term $(v\cdot\nabla)v\simeq(U\cdot\nabla)v\sim \mu\Delta\bv$ cannot be neglected \cite{lamb}}.
Thus defining the swimming efficiency as $\delta=(\bF\cdot\bU)/{\cal P}=
{\cal R} U^2/{\cal P}$ makes no sense in the present 2-D context in which $\bF,\:{\cal R}$ are not defined.
This may be considered as a mere issue of normalization.
We therefore use here an alternative definition of swimming efficiency\cite{AGK04} where
%Therefore, the optimal propulsion efficiency of the 2D microsimmer yields\footnote{The definition of the swimming efficiency $\delta$,
%differs from the one used in 3D, which due to the Stokes paradox, makes no sense in 2D.}
\be
\delta^{\star}=\frac{4\pi\mu U^2}{\mathcal{P}}={(1+\alpha)^2\over2\alpha}\log\left({1+\alpha\over 1-\alpha}\right).
\label{ddd}
\ee
In the slender limit, $\alpha\rightarrow1$, or, ${a\over b}\equiv {1-\alpha\over1+\alpha}=\varepsilon\rightarrow0$, as the ellipse degenerates into a needle, the efficiency grows logarithmically unbounded as
\[
\delta^{\star}\simeq 2\log(1/\varepsilon).
\]
It may be of interest to note that truncating our expansion to include any finite number of
modes would lead to $B(x)$ which is not only polynomial in $x$ but also algebraic in
$\alpha$. This then implies that the (truncated) efficiency $\delta^\star\propto B(0)$
would be algebraic in $\alpha$ implying that  \eqref{ddd} must be modified to a bounded
expression. Thus (in contrast to the 3-D case) one cannot obtain the correct asymptotic
efficiency without retaining all the modes.

The optimal boundary velocity may be found explicitly as
\ba
w(\theta)={\rd z\over \rd \theta}\sum u_k\sin(k\theta)={\rd z\over \rd\theta} \sum b_k\:\frac{(\re^{\ri (2k+1)\theta}-\re^{-\ri(2k+1)\theta})}{2\ri} \nonumber \\
=u_s{\rd z\over \rd \theta} \rIm\left\{\re^{\ri\theta}B\left(\re^{2\ri\theta}\right)\right\}. \qquad \qquad \qquad \nonumber
\ea
Using the explicit expression we have for $B(x)$ and ${\rd z\over \rd\theta}$ we find
the absolute value of $w$ is given by
\[
|w|=u_s{1+\alpha^2\over\sqrt{\alpha}}\log\left|{\sqrt{\alpha-\re^{2\ri\theta}}+ \sqrt{1/\alpha-\re^{2\ri\theta}}\over\sqrt{1/\alpha-\alpha}}\right|, \]
while its direction is known to be tangential. Thus, in the limit $\varepsilon \rightarrow 0$
one finds (provided $\varepsilon\ll\theta,\;|\pi-\theta|$) that
\[ |w/U| \simeq 1+{\log{|2\sin\theta|}\over\log{1/\varepsilon}},
%$+\mathcal{O}\left({\varepsilon\over\log\varepsilon}\right)
\] and the boundary velocity approaches the constant $U$ though only at a logarithmic rate.

\subsection{Concluding remarks}
In this paper we examined the propulsion of elongated microswimmer by virtue of the continuous surface treadmilling. As the slenderness increases, the hydrodynamic disturbance created by the surface motion diminishes, i.e. the microbot is propelled forward with the velocity of the surface treadmilling, while surface, except the near vicinity of the poles, remains stationary in the laboratory frame. As a result of that, the `cigar-shaped' treadmiler is self-propelled throughout almost quiescent fluid yielding very low viscous dissipation. The calculation of optimal hydrodynamic efficiency of the 3-D and the 2-D microswimmers reveals that the proposed swimming technique is not only superior to various motility mechanisms considered in the past, but also perform much better than dragging under the action of an external force.

\begin{acknowledgments}
This work was partially supported by Israel Science Foundation and the EU grant HPRN-CT-2002-00277 (to J.E.A.) and by the Fund of Promotion of Research at the Technion (to J.E.A. and A.M.L.).
\end{acknowledgments}

\end{article}

\begin{thebibliography}{99}

\bibitem{purcell} Purcell, E. M., (1977) \emph{Am. J. Phys.} \textbf{45}, 3-11.

\bibitem{berg} Berg, H. C., (2000) \emph{Phys. Today} \textbf{54}, 24-29.

\bibitem{PCC01} Pantaloni, D., Le Clainche, C., and Carlier, M. F. (2001) \emph{Science} \textbf{292}, 1502–1506.

\bibitem{TP89} Tilney, L.G., and Portnoy, D.A. (1989) \emph{J. Cell Biol.} \textbf{109}, 1597–1608.

%\bibitem{Dreyfus} Dreyfus, R, et al. (2005) \textit{Nature} \textbf{437}:...-....

%\bibitem{Kosa06} Kosa, G, (2006) Technion DSc Thesis.

%\bibitem{YLH} Yu, TS, Lauga, E and Hosoi, AE \texttt{cond-mat/0606527} and references therein.

%\bibitem{POS04} Plastino, J, Olivier, S.  and Sykes, C. (2004) \textit{Curr Biol} {\bf 14}: 1766-... .

\bibitem{CFOT99} Cameron, L. A., Footer, M. J., van Oudenaarden, A., and Theriot, J. A. (1999) \emph{Proc. Natl. Acad. Sci. USA} \textbf{96}, 4908–4913.

\bibitem{Anne02} Bernheim-Groswasser, A., Wiesner, S., Golsteyn, R. M., Carlier, M. F. and Sykes, C. (2002) \textit{Nature} \textbf{417}, 308–311.

\bibitem{UCASO03} Upadhyaya, A., Chabot, J.R., Andreeva, A., Samadani, A., and van Oudenaarden, A. (2003) \emph{Proc. Natl. Acad. Sci. USA} \textbf{100}, 4521–4526.

\bibitem{Leshansky06} Leshansky A. M. (2006) \textit{Phys Rev E} \textbf{74}, 012901-4.

\bibitem{DHP94} Dassios, G., Hadjinicolaou, M. and Payatakes, A. C. (1994) \textit{Quart. Appl. Math.} \textbf{52}, 157-191.

\bibitem{Zlat99} Zlatanovski, T. (1999) \textit{Q. J. Mech. Appl. Math.} {\bf 52}: 111-126.

\bibitem{HB64} Happel, J. and  Brenner H. (1965) in \textit{Low Reynolds Number Hydrodynamics} (Prentice-Hall, New Jersey).

\bibitem{KK91} Kim, S.  and Karrila, S. J. (1991)  in \textit{Microhydrodynamics} (Butterworth--Heinemann, Boston).

\bibitem{SS96} Stone, H and Samuel, E. M. (1996)  \textit{Phys. Rev. Lett.} \textbf{77}, 4102-4104.

\bibitem{SW89} Shapere, A. and  Wilczek, F. (1989) \textit{J. Fluid Mech.} {\bf 198}, 557-585.

\bibitem{ESBM96} Ehlers K. M., Samuel A., Berg, H. C. and Montgomery, R. (1996) \textit{Proc. Natl. Acad. Sci. USA} {\bf 93}, 8340-8343.

\bibitem{percel97} Purcell, E. M. (1997) \textit{Proc. Natl. Acad. Sci. USA}, \textbf{94}, 11307–11311.

\bibitem{WG98} Wiggins, C. H. and Goldstein, R. E. (1998) \textit{Phys. Rev. Lett.}, \textbf{80}, 3879–3882.

\bibitem {BKS03} Becker, L. E.,  Koehler, S. A. and Stone, H. A. (2003) \textit{J. Fluid Mech.}, \textbf{490}, 15–35.

\bibitem{AGK04} Avron, J. E., Gat, O. and Kenneth, O. (2004) \textit{Phys. Rev. Lett.} {\bf 93}, 186001-4.

%\bibitem{NG05} Najafi, A and Golestanian, R (2005) \textit{J Phys: Condens Matter} {\bf 17}: S1203-S1208.

%\bibitem{AKO05} Avron, JE, Kenneth, O and Oaknin, DH, \textit{ New J Physics}, {\bf 7}: 234-.

\bibitem{Langlois64} Langlois, W. E. (1964) in \emph{Slow Viscous Flow} (Macmillan, New York)

\bibitem{Rich68} Richardson, S. (1968), \textit{J. Fluid Mech.} {\bf 33}, 476-493.

%\bibitem{Antanovskii96} Antanovskii, LK (1996), \textit{J. Fluid Mech.} {\bf 327}, 325-341.

\bibitem{JM92} Jeong, J.-T. and Moffatt, H.K (1992), \textit{J. Fluid Mech.} {\bf 241}, 1-22.

\bibitem{lamb} Lamb, H. (1932) in \textit{Hydrodynamics} (Dover, New York).%, Art. 343.

\end{thebibliography}
\end{document}